\documentclass[prb,twoside,twocolumn,10pt,floatfix,showpacs,citeautoscript,superscriptaddress]{revtex4-2}

\usepackage[version=4]{mhchem}
\usepackage{amsmath}
\usepackage{amssymb}
\usepackage{booktabs}
\usepackage{glossaries}
\usepackage{graphicx}
\usepackage{tabularx}
\usepackage{siunitx}
\usepackage{hyperref}

\newcommand{\Title}{
    Quantitative predictions of the thermal conductivity in transition metal dichalcogenides: The impact of point defects in \texorpdfstring{\ce{MoS2}}{MoS2} and \texorpdfstring{\ce{WS2}}{WS2} monolayers
}

\hypersetup{
    pdfauthor={S. Mahendran , J. Carrete, A. Isacsson, G. K. H. Madsen, and P. Erhart},
    pdftitle={\Title},
    pdffitwindow=false,
    pdfstartview={FitH},
    pdfnewwindow=true,
    colorlinks=true,
    linkcolor=black,
    citecolor=black,
    filecolor=black,
    urlcolor=black,
}

\newcommand{\addchalmers}{Department of Physics, Chalmers University of Technology, SE-41296, Gothenburg, Sweden}
\newcommand{\addvienna}{Institute of Materials Chemistry, TU Wien, A-1060 Vienna, Austria}
\newcommand{\addzaragoza}{Instituto de Nanociencia y Materiales de Aragón (INMA), CSIC-Universidad de Zaragoza, E-50009 Zaragoza, Spain}

\graphicspath{{./figures/}}

\setacronymstyle{long-short}
\newacronym{bte}{BTE}{Boltzmann transport equation}
\newacronym{cx}{vdW-DF-cx}{van-der-Waals density functional with consistent exchange}
\newacronym{dft}{DFT}{density functional theory}
\newacronym{fc}{FC}{force constant}
\newacronym{fcp}{FCP}{force constant potential}
\newacronym{ltc}{LTC}{lattice thermal conductivity}
\newacronym{md}{MD}{molecular dynamics}
\newacronym{mfp}{MFP}{mean free path}
\newacronym{rmse}{RMSE}{root-mean-square error}
\newacronym{rta}{RTA}{relaxation time approximation}
\newacronym{sm}{SM}{Supplemental Material}
\newacronym{stem}{STEM}{scanning transmission electron microscopy}
\newacronym{stm}{STM}{scanning tunneling microscopy}
\newacronym{tmd}{TMD}{transition metal dichalcogenide}
\newacronym{xc}{XC}{exchange-correlation}

\renewcommand{\vec}[1]{\boldsymbol{#1}}
\renewcommand{\matrix}[1]{\mathbf{#1}}

\renewcommand{\Im}{\operatorname{Im}}
\newcommand{\qvec}{\boldsymbol{q}}

\DeclareSIUnit{\angstrom}{\text{Å}}
\DeclareSIUnit{\atom}{\text{atom}}
\DeclareSIUnit{\formulaunit}{\text{f.u.}}

\begin{document}

\title{\Title}

\author{Srinivisan Mahendran}
\affiliation{\addchalmers}
\author{Jesús Carrete}
\affiliation{\addzaragoza}
\affiliation{\addvienna}
\author{Andreas Isacsson}
\affiliation{\addchalmers}
\author{Georg K. H. Madsen}
\affiliation{\addvienna}
\author{Paul Erhart}
\email{erhart@chalmers.se}
\affiliation{\addchalmers}

\begin{abstract}
Transition metal dichalcogenides are investigated for various applications at the nanoscale thanks to their unique combination of properties and dimensionality.
For many of the anticipated applications, heat conduction plays an important role.
At the same time, these materials often contain relatively large amounts of point defects.
Here, we provide a systematic analysis of the impact of intrinsic and selected extrinsic defects on the lattice thermal conductivity of \ce{MoS2} and \ce{WS2} monolayers.
We combine Boltzmann transport theory and the Green's function-based $\boldsymbol{T}$-matrix approach for the calculation of scattering rates.
The force constants for the defect configurations are obtained from density functional theory calculations via a regression approach, which allows us to sample a rather large number of defects at a moderate computational cost and to systematically enforce both the translational and rotational acoustic sum rules.
The calculated lattice thermal conductivity is in quantitative agreement with experimental data for heat transport and defect concentrations for both \ce{MoS2} and \ce{WS2}.
Crucially, this demonstrates that the strong deviation from a $1/T$-temperature dependence of the lattice thermal conductivity observed experimentally, can be fully explained by the presence of point defects.
We furthermore predict the scattering strengths of the intrinsic defects to decrease in the sequence $\ce{V_{Mo}} \approx \mathrm{V}_\mathrm{2S}^{=} > \mathrm{V}_\mathrm{2S}^{\perp} > \ce{V_{S}} > \ce{S_{ad}}$ in both materials, while the scattering rates for the extrinsic (adatom) defects decrease with increasing mass such that $\ce{Li_{ad}} > \ce{Na_{ad}} > \ce{K_{ad}}$.
Compared to earlier work, we find that both intrinsic and extrinsic adatoms are relatively weak scatterers.
We attribute this difference to the treatment of the translational and rotational acoustic sum rules, which if not enforced can lead to spurious contributions in the zero-frequency limit.
\end{abstract}

\maketitle

\section{Introduction}

\Glspl{tmd} have emerged as fascinating materials in the fields of nanotechnology, electronics, and optoelectronics due to their quasi-two-dimensional (2D) structure and associated properties.
Notable examples are molybdenum disulphide (\ce{MoS2}) and tungsten disulphide (\ce{WS2}).
These 2D semiconducting compounds possess unique electronic, optical, and mechanical characteristics, distinct from those of their bulk counterparts.
The distinctive characteristics arise from their ultrathin atomic layers, which makes them promising candidates for the development of next-generation nanodevices, including transistors, sensors, photodetectors, and catalysts \cite{LinCarKah16, HonJinYua17, DinPenZho19, ZhoZhaSon20, LiaZhaZha21}.

The ability to efficiently conduct heat or, reciprocally, to impede its transport, can greatly impact the reliability, power dissipation, and thermal management of these materials in various applications, ranging from microelectronics to energy conversion systems \cite{Zheng_2021}.
It is therefore important to quantify the thermal conductivity and specifically the \gls{ltc} in these materials \cite{PisJacGaa16, JiaQiaGu17} and understand the limiting factors.
In this context, calculations can be tremendously useful, in particular when rooted in an ab-initio framework.
Yet, such calculations based on Boltzmann transport theory or Green-Kubo relations typically overestimate the \gls{ltc} and yield a stronger temperature dependence than observed experimentally\cite{LinErh16, PenZhaSha16}.
The deviation can be accounted for by invoking a heuristic, semi-empirical boundary scattering term, which obfuscates the underlying reason for the discrepancy.

For fully periodic, three-dimensional materials such as \ce{SiC} \cite{KatCarDon17} and \ce{GaAs} \cite{KunOttCar19}, it has been shown that the inclusion of point defect scattering is crucial for obtaining quantitative agreement with experimental measurements of the \gls{ltc}.
Thanks to the two-dimensional nature of \glspl{tmd} such as \ce{MoS2} and \ce{WS2}, it is possible to image defects directly, e.g., via \gls{stm} or \gls{stem}, which points to the S vacancy as the most abundant intrinsic defect \cite{LuLiMao14, HonHuPro15, CarWanFuj17, SonJooNeu17, TraNieBob22, LeeJeoKim22}.
These investigations have also shown the defect population to be sensitive to the synthesis conditions \cite{HonHuPro15} and to exhibit considerable spatial variations \cite{McDAddBui14, VanMagPet16, CarWanFuj17}.

While the majority of \gls{ltc} research on \glspl{tmd} so far has focused on phonon-phonon and isotope scattering in computational analyses, a few studies have also considered scattering by point defects and impurities \cite{DinPeiJia15, PenNinZha16, PolPanBer20, GabSurFar20}.
While these investigations clearly demonstrated the potential impact of defects, they are limited in scope, either because of the reliance on semi-empirical ingredients or because of the limited number of defect types considered (in particular extrinsic defects).
%A systematic, quantitative analysis with a connection to experimental data is still lacking.

Methods based on the Boltzmann transport formalism afford very detailed insight into phonon scattering, flexibility when it comes to defect concentrations and, when combined with ab-initio theory, proven predictive value.
However, in the case of quasi-2D systems very careful attention must be paid to the post-processing of the \emph{ab-initio} calculations to remove the effects of boundary conditions that violate the symmetries of free space \cite{CarLiLin16}.
With that in mind, here we provide a physically founded, systematic analysis of the impact of intrinsic and selected extrinsic defects on the \gls{ltc} of \ce{MoS2} and \ce{WS2} monolayers.
This allows us to provide a quantitative description of the measured \gls{ltc} without resorting to semi-empirical models and/or parameters.

\section{Methodology}
\label{sect:methodology}

\subsection{Lattice thermal conductivity}
\label{sect:lattice-thermal-conductivity}

The phononic contribution to the \gls{ltc} can be calculated by solving the \gls{bte}.
In this work the \gls{bte} is solved under the \gls{rta}, which yields the following expression for the \gls{ltc}
\begin{align}
    \kappa(T) = \frac{1}{2\Omega} \sum_{j\qvec}
    \vec{\lambda}_{j\qvec}(T) \otimes \vec{v}_{j\qvec} c_{j\qvec}(T).
    \label{eq:lattice-thermal-conductivity}
\end{align}
Here, $\Omega$ is the volume of unit cell while $\vec{\lambda}_{jq}$, $\vec{\upsilon}_{j\qvec}$, and $c_{j\qvec}$ are the phonon \gls{mfp} (interpreted as a vector), group velocity, and specific heat capacity of mode $j$ with momentum vector $\qvec$, respectively.
The \gls{mfp} $\vec{\lambda}_{jq} = \tau_{jq}(T) \vec{\upsilon}_{jq}$  is proportional to the relaxation time $\tau_{j\qvec}$.
In this work the inverse of the relaxation time, i.e., the total scattering rate  $\tau^{-1}_{j\qvec}$, is calculated using Matthiesen's rule by adding the contributions from phonon-phonon (ph-ph), isotope (iso), and defect scattering (def): 
\begin{align*}
    \tau^{-1}_{j\qvec} = \tau^{-1}_{\text{ph-ph},j\qvec} + \tau^{-1}_{\text{iso},j\qvec} + \sum_s^\text{defects} \tau^{-1}_{\text{def},s,j\qvec}.
\end{align*}
In this work the phonon-phonon scattering rate $\tau^{-1}_{\text{ph-ph},j\qvec}$, associated with the anharmonicity of the material is evaluated considering three-phonon processes.
The $\tau^{-1}_{\text{iso},j\qvec}$ term represents isotopic mass disorder and gives rise to a temperature-independent scattering rate.
The calculation of the defect scattering rates is described in the next section.
For comparison we also consider a model for boundary scattering given by
\begin{equation}
\tau^{-1}_{\text{b},j\qvec}=v_{j\qvec}/L,
\end{equation}
where $L$ is the characteristic length of the structural homogeneity of the material.

\subsection{Phonon scattering by defects}
\label{sect:phonon-scattering-by-defects}

The defect scattering rates can be obtained by using the optical theorem,
\begin{align*}
    \tau_{\text{def},j\qvec}^{-1} =
    - \rho_\text{def} \Omega         \frac{1}{\omega_{j\qvec}}
    \Im \{ \langle j'\qvec'|\matrix{T}|j\qvec\rangle \}.
\end{align*}
Here, $\rho_\text{def}$ is the volumetric defect concentration and $\omega_{j\qvec}$ is the angular phonon frequency.
The $\matrix{T}$-matrix links the phonon wave functions of the ideal and defect-laden systems through the relation
\begin{align*}
    \matrix{T} = (\matrix{I}-\matrix{V}g^{+})^{-1}\matrix{V},
\end{align*}
where $g^+$ is the retarded Green's function of the ideal structure while $\matrix{V}$ represents the perturbation connecting the ideal and the defect-laden systems, which can be decomposed as
\begin{align*}
    \matrix{V} = \matrix{V}_K + \matrix{V}_M.
\end{align*}
Here, the diagonal matrix $\matrix{V}_M$ describes changes in mass, with elements
\begin{align*}
    V_{M,ij} = \begin{cases}
        - \frac{M_i^\text{defect} - M_i^\text{pristine}}{M_i^\text{pristine}} \omega^2
        & i = j\\
        0 & i \neq j
    \end{cases},
\end{align*}
while $\matrix{V}_K$ captures changes in the \glspl{fc} and is given by
\begin{align*}
    V_{K,i\alpha j\beta} =
    \frac{K_{i\alpha j\beta}^\text{defect} - K_{i\alpha j\beta}^\text{pristine}
    }{\sqrt{M_i^\text{pristine} M_j^\text{pristine}}}.
\end{align*}
Here, $K_{i\alpha j\beta}$ is the \gls{fc} matrix where $i$ and $j$ are atomic indices while $\alpha$ and $\beta$ denote Cartesian indices.

In the cases of vacancies and adatoms, the creation of the defect can be conceptualized as the connection or disconnection of a subset of atoms to/from the bulk system.
The corresponding \glspl{fc} are zero in the pristine system for adatoms, and in the defect-laden system for vacancies, but the atoms themselves are not replaced.
The value of the mass perturbation becomes irrelevant, and can be taken as zero.
To successfully implement this idea for vacancies, the Green's function computed for the pristine system has to be augmented with a block corresponding to a free atom.

We use this Green's-function-based $\matrix{T}$-matrix approach as implemented in the \textsc{almaBTE} package \cite{CarVerKat17}, where the linear tetrahedron method \cite{Tetrahedron, Wang_PRB17} is used to calculate the Green's function of the ideal structure for each value of the incident phonon frequency.
This software package has previously been successful in computing thermal transport properties of solids in the presence of varying concentrations of defects, achieving good agreement with experiment \cite{KatCarDon17, KunOttCar19}.

\subsection{Computational details}
\label{sect:computational-details}

All structures and \glspl{fc} were obtained from \gls{dft} calculations that were carried out using the projector augmented-wave method \cite{Blo94} as implemented in the Vienna Ab-initio Simulation Package (VASP) \cite{KreHaf93, KreFur96}.
The exchange-correlation contribution was represented using the \gls{cx} method \cite{DioRydSch04, BerHyl2014}, which has been previously shown to provide an excellent description for \glspl{tmd} \cite{ErhHylLin15, LinErh16}.
All calculations were carried out using a plane-wave energy cutoff of \qty{260}{\electronvolt} and Gaussian smearing with a width of \qty{0.1}{\electronvolt}.
Projection operators were evaluated in reciprocal space and a finer support grid was employed during the calculation of the forces to improve the numerical accuracy of the latter.
For the primitive cells the Brillouin zone was sampled using a $\Gamma$-centered \numproduct{12x12x1} grid.
For the representation of defect structures as well as the calculation of \glspl{fc}, we used supercells comprising \numproduct{8x8x1} primitive cells (\num[separate-uncertainty = true]{192+-1} atoms); their Brillouin zone was sampled using a $\Gamma$-centered \numproduct{2x2x1} grid.
A vacuum layer of at least \qty{27}{\angstrom} was introduced along the axis perpendicular to the monolayer to avoid interactions between periodic images of these quasi-2D systems.
However, all values of the thermal conductivities are given for conventional thicknesses of \qty{6.15}{\angstrom} for \ce{MoS2} and \qty{6.18}{\angstrom} for \ce{WS2} \cite{LinErh16}.

The large defect-laden supercells, combined with their low symmetry, make it computationally expensive to evaluate the \glspl{fc} through systematic enumeration of the displacements.
For example, calculation of the \glspl{fc} for a sulphur vacancy would require \num{1146} separate \gls{dft} calculations using a direct finite-displacement approach.
We therefore resort to a regression approach \cite{EsfSto08, EriFraErh19} based on recursive feature elimination via \textsc{scikit-learn} \cite{scikit-learn} and the \textsc{hiphive} package \cite{EriFraErh19}.
This approach allows one to reduce the number of \gls{dft} calculations by one to two orders of magnitude depending on cell size \cite{FraEriErh20}.

To obtain the second-order \glspl{fc} for the different defect configurations, we generated structures with random displacements sampled from a Gaussian distribution with a standard deviation of \qty{0.01}{\angstrom}.
We then computed the forces for these reference structures via \gls{dft} (see above) and used them to construct \gls{fc} models, splitting them into a training set (75\% of the available data) and a validation set (25\% of the data).
To optimize the quality of the \gls{fc} extraction, we carried out convergence studies with respect to the cutoff imposed on pair interactions (which sets the number of free parameters) and the number of structures included in the training procedure.
The \gls{fc} models also include a small number of third-order terms according to a triplet cutoff of \qty{3}{\angstrom}, as this has been found to improve the quality of the second-order \glspl{fc} by avoiding the spurious inclusion of small anharmonic contributions \cite{FraEriErh20}.

\begin{figure}
    \centering
    \includegraphics{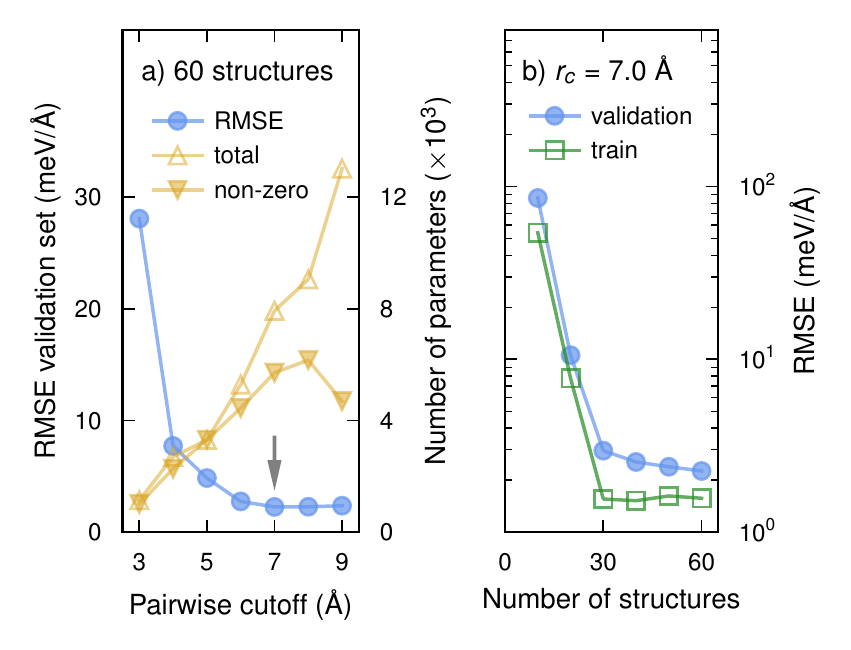}
    \caption{
        Extraction of force constants for defects.
        (a) Convergence of the root-mean-square error (RMSE) with the cutoff for pair interactions using a fixed number of reference structures for the sulphur vacancy (\ce{V_{S}}) in \ce{MoS2}.
        As the cutoff increases the total number of parameters increases as well, yet the number of non-zero parameters (as well as the RMSE) settle for a cutoff of \num{7}{\AA} and larger.
        (b) Convergence with the number of reference structures using a pairwise cutoff of \num{7}{\AA}.
    }
    \label{fig:force-constant-convergence}
\end{figure}

Using the sulphur vacancy (\ce{V_{S}}) in \ce{MoS2} as an illustrating examples, it can be seen that, with a fixed reference set size comprising \num{60} structures, the \gls{rmse} over the validation set quickly drops with increasing pair cutoff and then levels off at \qty{3}{\milli\electronvolt\per\angstrom} for a cutoff of \qty{7}{\angstrom} (\autoref{fig:force-constant-convergence}a).
Increasing the cutoff further does not yield a further reduction and the number of non-zero parameters remains approximately at the same level.

Next, we demonstrate the convergence with respect to the number of structures included in the reference set for a fixed pairwise cutoff of \qty{7}{\angstrom} (\autoref{fig:force-constant-convergence}b).
The \gls{rmse} for the validation set is already very low at \num{30} structures, suggesting that one could reduce the number of configurations for which \gls{dft} calculations have to be carried out by a factor of approximately \num{50}.
Here, we used, however, a more conservative number of \num{60} configurations, which still leads to very substantial savings in computational effort.
The convergence analysis for other defects yields practically identical results, so we proceeded with a pairwise cutoff of \qty{7}{\angstrom} and reference sets of \num{60} structures for all defects and both materials (\autoref{stab:computational-parameters}).

To obtain the second-order \glspl{fc} \ce{MoS2} and \ce{WS2}, we generated rattled \num{20} structures for each material based on the primitive 3-atom unit cell with an average displacement amplitude of \qty{0.2}{\angstrom}.
The numerical errors associated with \gls{dft} calculations often lead to a non-quadratic dispersion of the lowest transverse acoustic mode, which is a hallmark of two-dimensional materials.
Such artifacts can have a dramatic effect on the thermal conductivity \cite{CarLiLin16}.
It is therefore important to not only impose the translational but also the rotational acoustic sum rules.
In the present work, all sum rules are efficiently enforced via regularization using the \textsc{hiphive} package \cite{EriFraErh19}, so that even sets of interatomic \glspl{fc} for the defect-laden structures satisfy the symmetries of free space.

The third-order force constants are only needed for the perfects system and were calculated using a systematic finite-displacement approach as implemented in the \texttt{thirdorder} script included with the \textsc{ShengBTE} package  \cite{ShengBTE}.
We included interactions up to the seventh nearest neighbors, which yield cutoffs of \qty{7.08}{\angstrom} and \qty{7.09}{\angstrom} for \ce{MoS2} and \ce{WS2}, respectively.

To calculate the thermal conductivity, we integrated over the Brillouin zone employing an \numproduct{83x83x1} regular $\Gamma$-centered grid.
To avoid introducing artifacts in the bands through the use of the linear tetrahedron method, the calculation of the Green's function is performed on a denser \numproduct{103x103x1} grid.
Note that \num{83} and \num{103} are coprime numbers, so the two grids do not share any points other than $\Gamma$.

\section{Results}

\begin{figure}
    \centering
    \includegraphics[trim=0 30 0 0,clip]{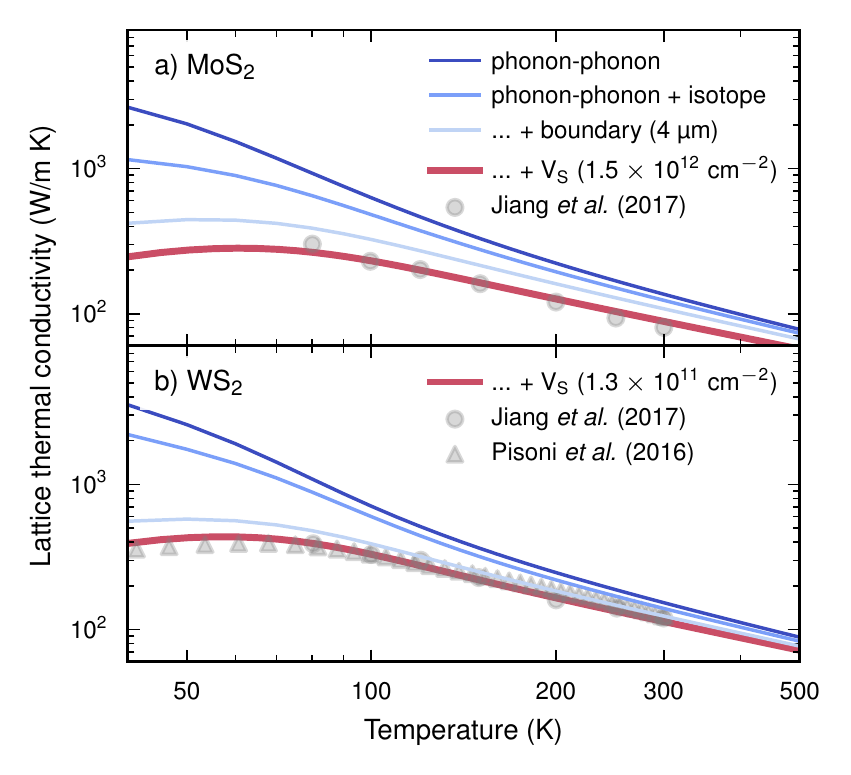}
    \includegraphics{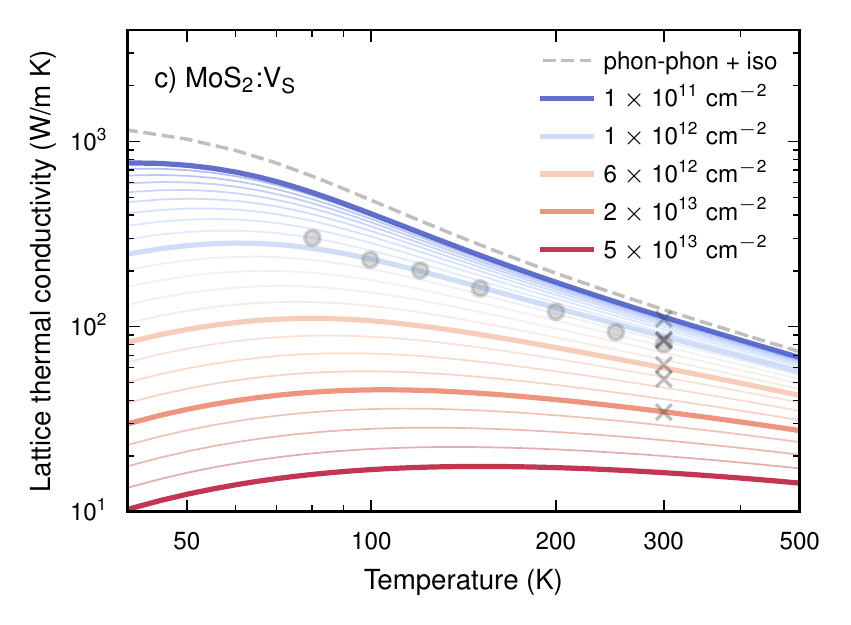}
    \caption{
        Lattice thermal conductivity of (a,c) \ce{MoS2} and (b) \ce{WS2} from calculations accounting for different scattering mechanisms including scattering by sulphur vacancies ($V_\text{S}$) at a fixed concentration (lines) and experiment (symbols).
        (c) Lattice thermal conductivity of \ce{MoS2} in the presence of S vacancies for a range of defect concentrations.
        Temperature-dependent experimental data from Jiang \textit{et al.} (circles; Ref.~\citenum{JiaQiaGu17}) and Pisoni \textit{et al.} (triangles; Ref.~\citenum{PisJacGaa16}).
        Additional experimental data from Refs.~\citenum{SahGauAhm13, LiuChoCah14, YanSimBer14, TauJudLap15, ZhaSunLi15} are shown by crosses in (c).
    }
    \label{fig:thermal-conductivity-vs-temperature}
\end{figure}

When considering phonon-phonon scattering only, one obtains a \gls{ltc} that notably overestimates the experimental data \cite{JiaQiaGu17} (\autoref{fig:thermal-conductivity-vs-temperature}a,b), in particular at lower temperatures; one also observes a more pronounced temperature dependence.
While including isotopic mass-variance scattering lowers the \gls{ltc} somewhat the prediction is still far from the experimentally measured values.

A semi-empirical way to account for the difference is to introduce a boundary scattering term, which introduces an intrinsic length scale that caps the mean free paths of the phonons.
This is illustrated here using the length parameter of \qty{4}{\micro\meter} used in Ref.~\citenum{LinErh16} (light-blue lines in \autoref{fig:thermal-conductivity-vs-temperature}a,b).
This value is considerably smaller than typical grain sizes in the samples used for comparison here, and while it enables a fit to the data it fails to provide deeper physical insight.

It is well established that point-defect concentrations in \glspl{tmd} tend to be rather high, with densities up to \qty{e13}{\per\centi\meter\squared} \cite{ZhoZouNaj13, LuLiMao14, McDAddBui14, HonHuPro15, VanMagPet16, SonJooNeu17, CarWanFuj17, RosChuMcC18, LiaZhaZha21, LeeJeoKim22, TraNieBob22}.
This raises the question of whether defects can quantitatively account for the gap between prediction and measurement, and if so how effective different defects with respect to phonon scattering.
To answer these questions, we computed the scattering rates for a selection of intrinsic and extrinsic defects that have been frequently observed experimentally \cite{HonHuPro15, PisJacGaa16, TraNieBob22, LeeJeoKim22}.
Among the intrinsic defects we included the Mo (\ce{V_{Mo}}), W (\ce{V_W}), and S monovacancies (\ce{V_S}), the in-plane ($\mathrm{V}_\mathrm{2S}^{=}$) and out-of-plane S divacancies ($\mathrm{V}_\mathrm{2S}^{\perp}$) as well as the S adatom (\ce{S_{ad}}).
According to \gls{dft} calculations these defects adopt a neutral charge state for most electron chemical potentials \cite{KomKra15, KieDurMil23}.
We therefore restricted ourselves to the neutral charge state.

Among the extrinsic defects we considered adatoms of Li, Na, and K in the case of \ce{MoS2} and of Na in \ce{WS2} as representatives of impurities introduced during exfoliation but also as prototypes for other impurities and dopants that commonly occupy adatom sites.
The three species also span a range of masses, which gives rise to spread in the frequencies of defect related modes.
Two different configurations were investigated:
In the $X_\text{ad1}$ geometry the $X$ atoms reside directly above Mo/W whereas in the $X_\text{ad2}$ configuration the adatom sits above an empty ``channel''.

While \Gls{stm} and \gls{stem} measurements enable measuring defect concentrations, this is still challenging, both due to the statistics involved and technical complexities such as concurrent beam damage in the case of electron microscopy \cite{KomKurLeh13}.
For \ce{MoS2} defect concentrations between \num{4e10} and \qty{5e13}{\per\centi\meter\squared} have been measured \cite{LuLiMao14, McDAddBui14, HonHuPro15, JeoLeeLy16, VanMagPet16, TraNieBob22}.
The data available for \ce{WS2} is sparser.
Using \gls{stem} a bulk defect concentration of \qty{3.3e13}{\per\centi\meter\squared} has been obtained \cite{CarWanFuj17}, while \gls{stm} measurements yielded a density of \num{2.3e10} for Mo and \qty{4.5e11}{\per\centi\meter\squared} for S vacancies \cite{RosChuMcC18}.
Given that electron irradiation is known to facilitate defect formation \cite{KomKurLeh13}, it is plausible that the \gls{stem} values could be overestimates.
The \gls{stm} data on the other hand indicate that defect concentrations in \ce{WS2} might be lower than in \ce{MoS2}, a suggestion that is supported by our analysis as will be discussed below.

\begin{figure}
    \centering
    \includegraphics{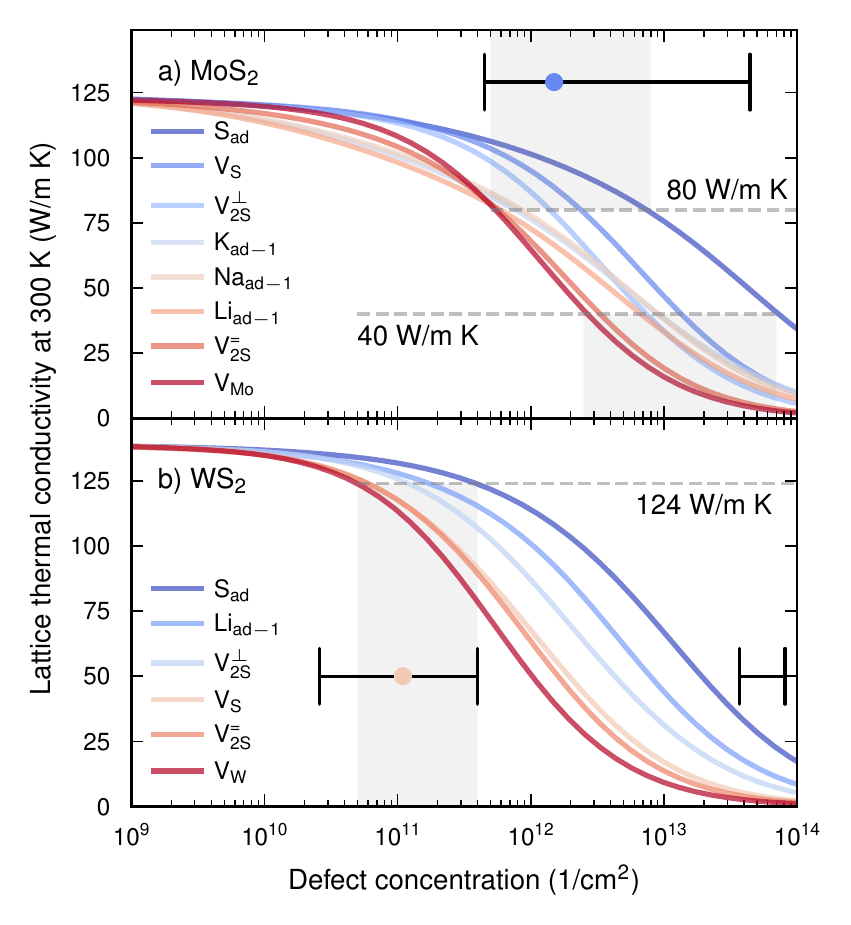}
    \caption{
        Lattice thermal conductivity at \qty{300}{\kelvin} of (a) \ce{MoS2} and (b) \ce{WS2} as a function of concentration for different defects.
        Gray rectangles indicate the concentration range that approximately aligns with typical experimental values for the \gls{ltc}.
        For \ce{MoS2} both an approximate upper (\qty{80}{\watt\per\kelvin\meter}) and lower value (\qty{40}{\watt\per\kelvin\meter}) are indicated.
        The horizontal bars represent the range of experimentally determined defect concentrations (see text for details).
        In the case of \ce{WS2} the data from Refs.~\citenum{CarWanFuj17} and \citenum{RosChuMcC18} are shown separately.
        The colored circles indicate the S vacancy concentrations used in \autoref{fig:thermal-conductivity-vs-temperature}a,b.
    }
    \label{fig:thermal-conductivity-vs-defect-concentration-MoS2}
\end{figure}

\begin{figure*}
    \centering
    \includegraphics{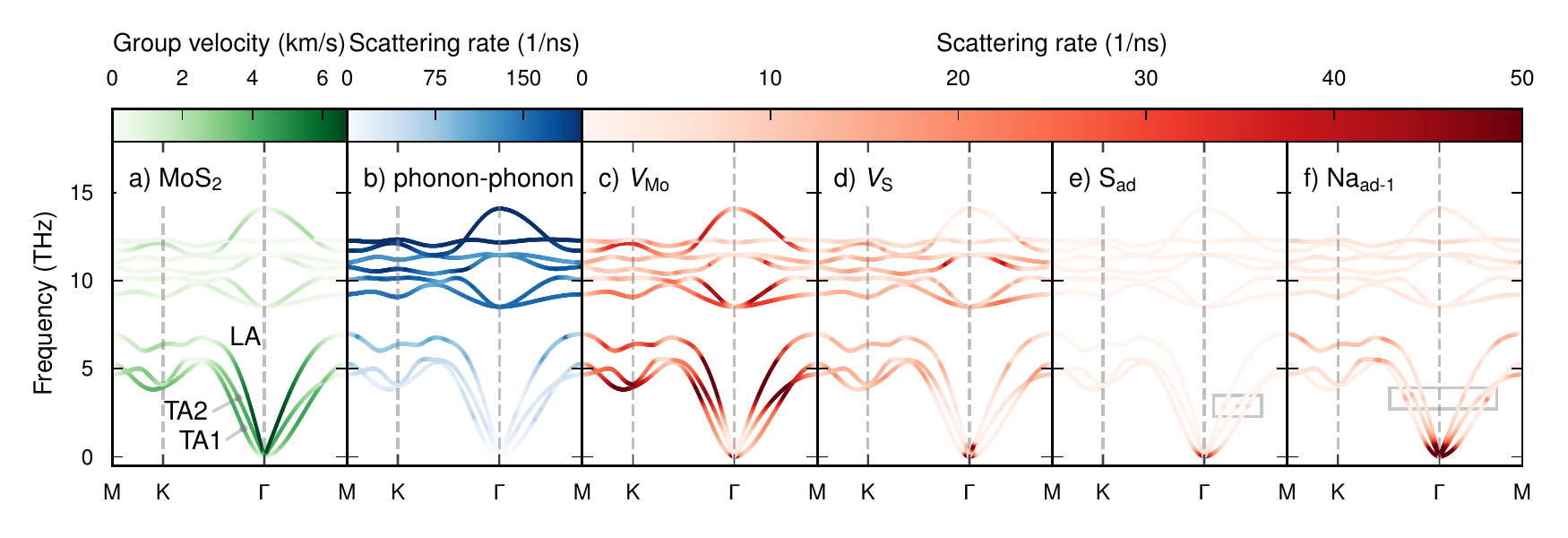}
    \caption{
        Phonon dispersion relations for \ce{MoS2} showing the (a) group velocities and (b) phonon-phonon scattering rates at \qty{300}{\kelvin} as well as the defect scattering rates for (c) $V_\text{S}$, (d) $V_\text{Mo}$, (e) $\text{S}_\text{ad}$, and (f) $\text{Na}_\text{ad-1}$ defects at a concentration of \num{1} defect per \num{e3} unit cells (approximately \qty{e11}{\per\centi\meter\squared}).
        We choose this convention to enable a more direct comparison among defect types, independent of the numbers of atoms of each type in the unit cell.
    }
    \label{fig:phonon-dispersion-with-scattering-rates-MoS2}
\end{figure*}

Since S vacancies are the most widely observed defect, we first focus our analysis on them.
Increasing the concentration of S vacancies (and point defects generally, see \gls{sm}) leads to a drop of the \gls{ltc} and more importantly a less pronounced temperature dependence (see \autoref{fig:thermal-conductivity-vs-temperature}c for \ce{MoS2} and \autoref{sfig:thermal-conductivity-WS2-S-vac} for \ce{WS2}).
This is expected since the rates of elastic scattering induced by defects are temperature independent, whereas inelastic three-phonon processes become more likely as temperature increases.
In \ce{MoS2} at room temperature the \gls{ltc} varies between \qty{135}{\watt\per\kelvin\per\meter} at a vacancy concentration of \qty{e10}{\per\centi\meter\squared} to \qty{9}{\watt\per\kelvin\per\meter} at \qty{e13}{\per\centi\meter\squared} (\autoref{fig:thermal-conductivity-vs-defect-concentration-MoS2}a).
This range includes experimental measurements of the \gls{ltc}, which span from \num{120} (Ref.~\citenum{LiuChoCah14}) to \qty{35}{\watt\per\kelvin\per\meter} (Ref.~\citenum{YanSimBer14}), which would be equivalent to defect densities of \num{7e10} and \qty{2e12}{\per\centi\meter\squared}, respectively.

Crucially, using typical defect densities from the middle of the experimentally observed range (see horizontal bars in \autoref{fig:thermal-conductivity-vs-defect-concentration-MoS2}) yields good agreement with experimental data over the entire temperature range.\footnote{
    Unfortunately, to the best of our knowledge, there are no studies for which defect concentrations and \gls{ltc} have been measured for the same sample.
}
For \ce{MoS2} and \ce{WS2} this suggests defect concentrations of approximately \num{1.5e12} and \qty{1.3e11}{\per\centi\meter\squared}, respectively.
These concentrations thus differ by one order of magnitude, in line with the differences observed experimentally between these two materials (see above).

Up to now we have considered only one type of defect but obviously defects differ in their scattering efficacy.
This can be illustrated by analyzing the variation of the \gls{ltc} with defect concentration at a fixed temperature (\qty{300}{\kelvin} in \autoref{fig:thermal-conductivity-vs-defect-concentration-MoS2}; see \autoref{sfig:thermal-conductivity-vs-defect-concentration-overview} for more temperatures).
In both \ce{MoS2} and \ce{WS2}, S adatoms are the weakest and metal vacancies (\ce{V_{Mo}}, \ce{V_W}) are the strongest scatterers.
However, as the concentrations needed to accomplish a certain \gls{ltc} reduction only vary by about one order of magnitude (as indicated by the gray rectangles in \autoref{fig:thermal-conductivity-vs-defect-concentration-MoS2}), the overall scattering efficiency is similar.

So far we have shown that inclusion of defect scattering when predicting the \gls{ltc} leads to near-quantitative agreement with experimental data without resorting to empirical parameters.
It is now instructive to investigate the microscopic mechanism by which defects scatter phonons.
Within the framework of Boltzmann transport theory the \gls{ltc} is predominantly determined by the phonon group velocities and lifetimes.

It is apparent from the phonon dispersion (\autoref{fig:phonon-dispersion-with-scattering-rates-MoS2}a for \ce{MoS2}; see also \autoref{sfig:phonon-dispersion-with-scattering-rates-WS2-defects} for \ce{WS2}) that the acoustic branches are the main contributors to the thermal conductivity given their more significant group velocities and lower frequencies.
As expected, the largest speed of sound is found in the longitudinal acoustic branch, followed by the transverse acoustic branches (labelled TA1 and TA2 in \autoref{fig:phonon-dispersion-with-scattering-rates-MoS2}a).
The optical branches are only weakly dispersive; not only does this lead to lower group velocities but also enables more opportunities for the conservation of energy in three-phonon processes, leading to stronger anharmonic scattering as seen in panel \autoref{fig:phonon-dispersion-with-scattering-rates-MoS2}b.

The overall magnitude of the phonon-defect rates follows the scattering efficiency observed via the \gls{ltc}, that is \ce{V_{Mo}} scatters more strongly than \ce{V_S}, which scatters more strongly than \ce{S_{ad}} (\autoref{fig:phonon-dispersion-with-scattering-rates-MoS2}c--e).
All defects lead to scattering across the entire dispersion, but the lowermost acoustic modes close to the center of the Brillouin zone as well as the higher-frequency range of the acoustic branches are the most notably affected.
The zero-frequency limit of the scattering rates deserves special mention since it deviates from the familiar Rayleigh power-law found in 3D systems \cite{Tamura}.
In fact, the elastic scattering rates do not vanish as $\omega\to 0$ in quasi-2D systems due to the nonzero density of states in that limit.

There are signatures of resonant scattering for the S adatom as well as for the extrinsic adatoms (\autoref{fig:phonon-dispersion-with-scattering-rates-MoS2}e,f and \autoref{sfig:phonon-dispersion-with-scattering-rates-WS2-defects}) in the form of maxima in the scattering rates along the acoustic branches around \qty{2.9}{\tera\hertz} (\ce{S_{ad}}) and \qty{3.3}{\tera\hertz} (\ce{Na_{ad-1}}; marked by rectangles in \autoref{fig:phonon-dispersion-with-scattering-rates-MoS2}e,f).
This effect is, however, less pronounced than in, e.g., B-doped \ce{SiC} \cite{KatCarDon17} or for DX-center defects in \ce{GaAs} \cite{KunOttCar19}, and does not reach the level required to create a ``superscatterer''.

\section{Discussion}

It is now instructive to compare our results and analysis with previous investigations.

Ding \textit{et al.} employed an empirical interatomic potential and nonequilibrium \gls{md} simulations with a thermal gradient to study the effect of Mo and S vacancies as well as S adatoms on the \gls{ltc} in \ce{MoS2} \cite{DinPeiJia15}.
They considered defect densities between \qty{1.2e12}{\per\centi\meter\squared} and \qty{1.2e13}{\per\centi\meter\squared}, at the upper end of the experimentally observed range (see below), and observed a reduction of the \gls{ltc} by 35 to 60\% at \qty{300}{\kelvin}.
% cross section of one formula unit MoS2: 8.58657797 Å^2 = 8.58657797e-16 cm^2
% 100% = 2 / 8.586577971e-16 / cm^2
% 0.05% = 0.5e-3 * 2 / 8.586577971e-16 = 1.2e12 / cm^2
% 0.5% = 0.5e-2 * 2 / 8.586577971e-16 = 1.2e13 / cm^2

Peng \textit{et al.}, on the other hand, used Boltzmann transport theory with third-order \glspl{fc} from \gls{dft} calculations to assess the impact of S mono and divacancies in \ce{MoS2} on the \gls{ltc} \cite{PenNinZha16}.
As a result of their approach, the defect concentration was fixed by the supercell size yielding defect densities of \qty{1.2e14}{\per\centi\meter\squared} and above.
They obtained a reduction for the in-plane \gls{ltc} by up to 75\% at \qty{300}{\kelvin}.
Moreover, this method treats scattering by those vacancies as inelastic, does so perturbatively, and assumes a perfectly periodic arrangement of the defects.
% 100% = 2 / 8.586577971e-16 / cm^2
% 5% = 0.5e-2 * 2 / 8.586577971e-16 = 1.2e14 / cm^2
% 10% = 1e-2 * 2 / 8.586577971e-16 = 2.4e14 / cm^2

Combining Boltzmann transport theory with the $\matrix{T}$-matrix approach, Polanco and co-authors characterized the dependence of the \gls{ltc} of \ce{MoS2} on both temperature and the concentration of S vacancies and adatoms \cite{PolPanBer20}.
They concluded that the spread in the measured \gls{ltc} is consistent with the experimentally observed variation in defect concentrations.
According to their calculations, adatoms are much more effective phonon scatterers than vacancies, owing mainly to the fact that only the phonon-defect scattering rates induced by the latter decay to zero at low frequencies.

Finally, using an empirical potential and equilibrium \gls{md} simulations, Gabourie \textit{et al.} mapped out the dependence of the \gls{ltc} at \qty{300}{\kelvin} on the concentration of S vacancies in both suspended and supported \ce{MoS2} \cite{GabSurFar20}.
The simulation approach enabled them to access more realistic defect densities between \num{e12} and \qty{5e13}{\per\centi\meter\squared}, and they found a 19\% reduction already at the lower end of the concentration range considered, arriving at a similar conclusion as Ref.~\citenum{PolPanBer20}.

Our results deviate quantitatively and qualitatively from both the \gls{md} simulations by Ding \textit{et al.}\cite{DinPeiJia15} and from the \gls{bte}-based predictions by Polanco \textit{et al.} \cite{PolPanBer20}
Both of those references suggest that S adatoms are significantly stronger scatterers than the S vacancy, and indeed that the former depresses the thermal conductivity to a much larger degree at a comparable concentration.
However, Ref.~\citenum{DinPeiJia15} relies on a semiempirical potential and can be expected to provide limited quantitative insight into configurations where atomic environments are very different from those of the bulk, as is the case of the S adatom.
Reference~\citenum{PolPanBer20}, on the other hand, employs a methodology much closer to our work and allows for a more detailed comparison.
This reveals that the phonon--S-adatom scattering rates reported by Polanco \textit{et al.} have a finite zero-frequency limit, but the rates they report for the vacancy vanish as $\omega\to 0$.
In the absence of a stated physical argument that could justify this contrasting behavior, the hypothesis of a numerical artifact cannot be ruled out.
Specifically, the enforcement of the rotational acoustic sum rules, necessary for both the \glspl{fc} of the pristine system and those of the defect-laden structures, can be a significant challenge, and as stated above the Green's function for the pristine system requires an extra postprocessing step to be able to deal with interstitial or adsorbed atoms.
Regarding the acoustic sum rules, we are confident in our systematic use of the extensively tested \textsc{hiphive} package \cite{EriFraErh19} for all structures.
Additionally, comparison among the results for the sequence of alkaline adatoms $\text{Li}_{\text{ad-1}}$, $\text{Na}_{\text{ad-1}}$, and $\text{K}_{\text{ad-1}}$ in \autoref{fig:thermal-conductivity-vs-defect-concentration-MoS2} serves as a qualitative test of our treatment of adatoms:
The effect of the structurally and chemically similar \ce{Li_{ad-1}} and \ce{Na_{ad-1}} defects on the thermal conductivity is very much alike, but with the heavier \ce{Na_{ad-1}} acting as a more intense scatterer, as expected.
This reasoning also holds for \ce{K_{ad-1}} in the high-concentration regime where elastic scattering dominates.
In this context, the effect of the sulphur adatom is perfectly reasonable for an element of its mass.

\section{Conclusions}

In this work we have quantified the impact of point defect scattering on the \gls{ltc} in \ce{MoS2} and \ce{WS2}.
When considering only phonon-phonon and isotope scattering, the computed \gls{ltc} significantly overestimates the experimental values, especially at low temperatures.

Defects, particularly sulfur vacancies, play a vital role in mediating phonon scattering in these \glspl{tmd}.
Our study suggests that the typically high point-defect concentrations in these materials, which may reach and exceed \qty{e13}{\per\centi\meter\squared}, can substantially influence the lattice thermal transport.
Calculations on a range of both intrinsic and extrinsic defects prevalent in these materials confirmed the S vacancy as the most dominant scattering center in terms of abundance and strength.
Importantly, using defect concentrations consistent with those reported in experimental studies brings our \gls{ltc} predictions in quantitative alignment with the measured data across the temperature range.
This demonstrates not only the importance of point defects for understanding the thermal conduction in these materials but suggests that variations in measured \gls{ltc} values can be related to differences in the point defect distributions.
Compared to earlier work, we find that both intrinsic and extrinsic adatoms are relatively weak scatterers, a difference that we attribute to the treatment of the translational and rotational acoustic sum rules.
In the present work, these were enforced, which removes spurious contributions in the zero-frequency limit that can otherwise affect the results.

In summary, our work underscores the importance of defect-mediated phonon scattering in governing the thermal transport properties of \ce{MoS2} and \ce{WS2}, and probably also in other transition metal dichalcogenides.
This deeper understanding of the underlying mechanisms provides valuable insights for tailoring the thermal properties of these materials, paving the way for their potential applications in next-generation electronic and thermal management devices.

\section*{Acknowledgements}

This work was funded by the Knut and Alice Wallenberg Foundation (2014.0226) and the Swedish Research Council (grant numbers 2017-06819, 2018-06482, 2019-03993, 2020-04935, 2021-05072).
We also acknowledge support by the Austrian Science Fund (FWF), doctoral college TU-DX (DOC 142-N).
The computations were enabled by resources provided by the National Academic Infrastructure for Supercomputing in Sweden (NAISS) and the Swedish National Infrastructure for Computing (SNIC) at C3SE, NSC, and PDC partially funded by the Swedish Research Council through grant agreements no. 2022-06725 and no. 2018-05973.
We gratefully acknowledge Daniel Lindroth's contribution during the early phase of this project.

\end{document}